\documentclass[aps,prl,twocolumn,superscriptaddress]{revtex4-1}

\usepackage{graphicx,graphics}
\usepackage{amsmath,amssymb}
\usepackage{epstopdf}

\newcommand{\nb}{\bar{n}}
\newcommand{\nbd}{\dot{\bar{n}}}
\newcommand{\Gd}{\Gamma_\downarrow}
\newcommand{\Gu}{\Gamma_\uparrow}

\begin{document}
\title{Brillouin Cooling}

\author{Matthew~Tomes}
\affiliation{Department of Electrical Engineering, University of Michigan, Ann Arbor, MI, 48109 USA}
\author{Florian~Marquardt}
\affiliation{Institut f\"ur Theoretische Physik, Universit\"at Erlangen-N\"urnberg, Staudtstrasse 7, D-91058 Erlangen, Germany}
\affiliation{Max Planck Institute for the Science of Light, G\"unther-Scharowsky-Strasse 1/Bau 24, D-91058 Erlangen, Germany}
\author{Gaurav~Bahl}
\affiliation{Department of Electrical Engineering, University of Michigan, Ann Arbor, MI, 48109 USA}
\author{Tal~Carmon}
\email{tcarmon@umich.edu}
\affiliation{Department of Electrical Engineering, University of Michigan, Ann Arbor, MI, 48109 USA}
%\maketitle

\begin{abstract}
We analyze how to exploit Brillouin scattering for the purpose of cooling opto-mechanical devices and present a quantum-mechanical theory for Brillouin cooling. Our analysis shows that significant cooling ratios can be obtained with standard experimental parameters. A further improvement of cooling efficiency is possible by increasing the dissipation of the optical anti-Stokes resonance.

\end{abstract}
\maketitle

\begin{figure}[tbp]
    \centering
    \includegraphics[width=.9\hsize]{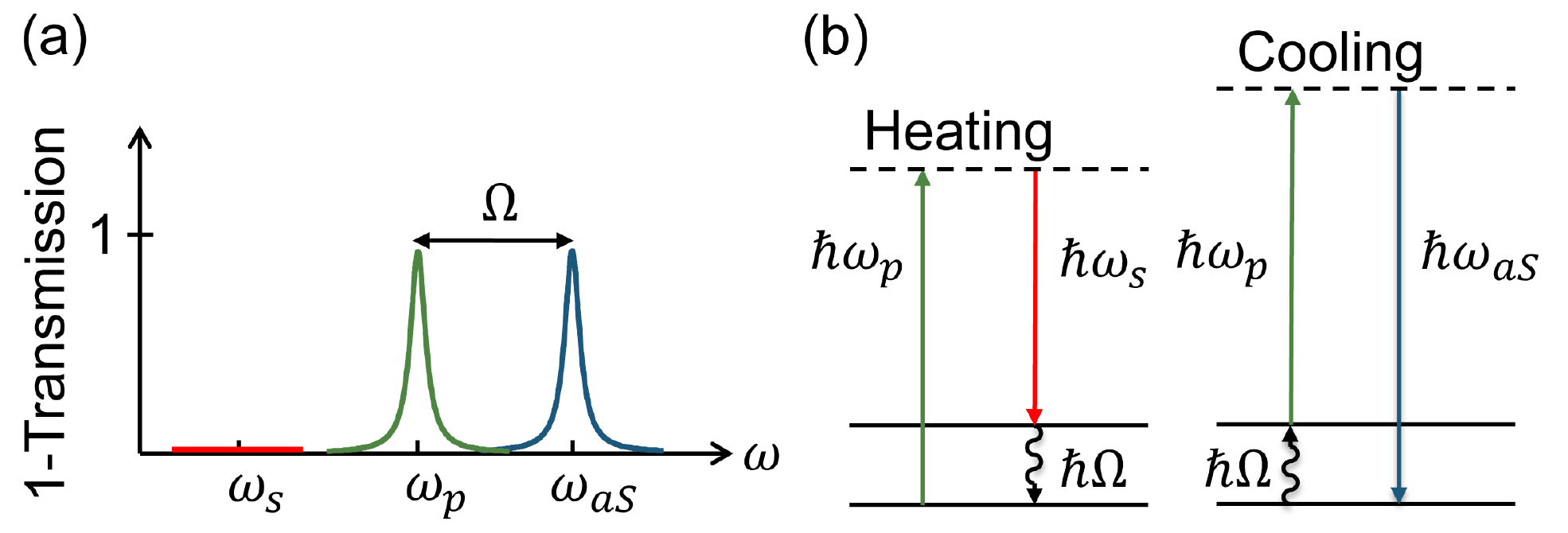}
    \caption{\label{figbrillgain}
    a) Sample resonator spectrum showing resonant enhancement possible at both pump and anti-Stokes frequencies, with attenuation at Stokes frequency. b) Energy diagram for Brillouin heating and cooling processes.}
\end{figure}

\emph{Introduction.---}A lesser known quality of Brillouin scattering is its ability to scatter photons in the anti-Stokes direction \cite{xboyd}, while cooling the corresponding mechanical vibration. The recent experimental observation of such Brillouin cooling \cite{xBahl2011b} suggests that a model is required to describe the potential of this new system for ground-state cooling \cite{Marquardt2007,WilsonRae2007}. In such light-sound interactions, light is scattered to both Stokes and anti-Stokes side bands, which heat and cool the system (Fig. \ref{figbrillgain}b).

In order to break the material heating-cooling symmetry, we study here a setup involving a resonator with an asymmetric resonance structure. Figure \ref{figbrillgain}a exemplifies our proposal to resonantly enhance the anti-Stokes process for cooling the mode, while at the same time off-resonantly attenuating the  Stokes process to prevent heating. Obtaining a resonator that is proper for Brillouin cooling is challenging. This is because two optical resonances that have almost the same optical frequency  but different propagation constants  are needed in order to conseerve both the energy and momentum which are given to light by the acoustical phonon. One type of cavity that allows such optical-resonance pairs is whispering-gallery mode resonators \cite{Carmon2008,Savchenkov2007}. In such resonators, the transverse (radial-polar) order of one mode can compensate for the frequency difference originating from the non-similar longitudinal (azimuthal) order of the other. This provides a pair of modes with different azimuthal wavevectors, but nearby frequencies, as experimentally observed via the resulting stationary interference pattern \cite{Carmon2008,Savchenkov2007}. The energy flow in Brillouin anti-Stokes cooling \cite{xBahl2011b} that is analyzed here, is opposite in respect to the Stokes excitation process \cite{Grudinin2009,Tomes2009,Bahl2011}. Additionally, cooling here is a spontaneous process.

\emph{Derivation of Optoacoustic Coupling Rate.---}First we derive the optoacoustic coupling rate $\Gamma_{opt}$ associated with Brillouin cooling. Consider a purely longitudinal density fluctuation that is propagating along L through a cross-sectional area $A$. A displacement field $u$ causes an elastic potential of $\frac{1}{2}AT(\partial_zu)^2$, where $T$ is the spring constant per area. Additionally, the kinetic energy will be $\frac{1}{2}A\rho\dot{u}^2$ where $\rho$ is the mass density. The interaction with the light field $E$ is given by $\frac{1}{2}A\gamma(\partial_zu)E^2$, where $\gamma=\rho\frac{\partial\epsilon}{\partial\rho}$ is the electrostrictive constant which relates a change in density to a change in permittivity. Thus our Hamiltonian is of the form:
    \begin{align}\label{eqHam}
    \hat{H}=\int^L_0dz\left\lbrace\frac{\hat{\pi}^2(z)}{2\rho A}+\frac{1}{2}AT(\partial_z\hat{u})^2+\frac{1}{2}A\gamma(\partial_z\hat{u})\hat{E}^2+...\right\rbrace~,
    \end{align}
where the omitted parts refer to the Hamiltonian of the electromagnetic field inside the medium. Here $\hat{\pi}(z)=A\rho\dot{\hat{u}}$ is the momentum density.

\emph{Quantization of the Electromagnetic Field and Sound Wave.---}For the electromagnetic field, consider a single polarization subjected to periodic boundaries. The electric field is of the form:
    \begin{align}\label{eqEkz}
    \hat{E}(z)=\sum_kE_k\left[\hat{a}_ke^{ikz}+h.c.\right]~,
    \end{align}
where $E_k$ is the zero-point amplitude of the electric field. Knowing the total energy of the electric and magnetic fields in free space, $E^2$ and $B^2$, to be $\frac{\hbar\omega_k}{4}$ per mode for the ground-state, we solve for the zero point fluctuation of the electric field,
    \begin{align}
    E_k&=\sqrt{(\hbar\omega_k)/(2AL\epsilon)}~.
    \end{align}
Similar to the electric field, the longitudinal sound wave will be quantized in the following form:
    \begin{align}\label{equkz}
    \hat{u}(z)=\sum_k u_k\left[\hat{b}_ke^{ikz}+h.c\right]~.
    \end{align}
Substituting Eq. \eqref{equkz} into Eq. \eqref{eqHam} and solving for the coefficient of $\hat{b}_k\hat{b}^\dagger_k$, we obtain the expressions:
    \begin{align}
    u_k&=\sqrt{(\hbar)/(2AL\rho\omega_k)}~,
%    \omega_k&=\vert k\vert\sqrt{T/\rho}=\vert k\vert v_s~,
    \end{align}
where $\omega_k=\vert k\vert v_s$ is the acoustic dispersion relation, and $v_s=\sqrt{T/\rho}$ is the speed of sound.

\emph{Acousto-optical interaction.---}
Having derived expressions for both $E_k$ and $u_k$ we turn our attention to the optomechanical interaction term. Substituting expressions for $E_k$ and $u_k$ from Eqs. \eqref{eqEkz} and \eqref{equkz} into \eqref{eqHam} and using the rotating wave approximation, the Hamiltonian simplifies to the following form:
    \begin{align}
    \notag &\hat{H}_{optmech}=\int^L_0dz(1/2)A\gamma(\partial_z\hat{u})\hat{E}^2(z)~,\\
    \notag &=\left(\frac{AL\gamma}{2}\right)\sum_{k,q}qu_qE_kE_{k+q}(\hat{b}_q+\hat{b}^\dagger_{-q})(\hat{a}^\dagger_{k+q}\hat{a}_{k}+\hat{a}_{k}\hat{a}^\dagger_{k+q})~,\\
    &=\hbar\sum_{k,q}g_{k,q}(\hat{b}_q+\hat{b}^\dagger_{-q})(\hat{a}^\dagger_{k+q}\hat{a}_{k}+\hat{a}_{k}\hat{a}^\dagger_{k+q})~,
    \end{align}
where $\hbar g_{k,q}=(1/2)AL\gamma qu_qE_kE_{k+q}$.

The pump mode (p) is coupled to another optical mode (aS) and an acoustical mode in which energy conservation ($\omega_p=\omega_{aS}-\Omega$) and momentum conservation ($k_p=k_{aS}-q$) are fulfilled \cite{Bahl2011}. Here, $q$ and $\Omega$ represent the wavevector and frequency of the acoustical mode. Conservation of both quantities is possible in the kind of experimental setup studied here due to high order transverse optical modes \cite{Carmon2008,Savchenkov2007}. Taking into account the momentum and energy conservation, the relevant coupling term in $\hat{H}$ for this process will be:
    \begin{align}
    \hbar g_{k,q}(\hat{b}_q\hat{a}^\dagger_{k+q}\hat{a}_k+h.c.)~.
    \end{align}
%
%%%%%%%%%%%%%
% AMPLITUDE EQUATIONS %
%%%%%%%%%%%%%

\emph{Solving the amplitude equations.---}
As long as Brillouin cooling does not change the optical mode populations significantly, we can describe the interaction using coupled amplitude equations for the light field and the vibrational mode. We can write a Hamiltonian for the laser-driven mode $\hat{a_1}$, the anti-Stokes mode $\hat{a_2}$, and the phonon mode $\hat{b}$ of the form (written in a frame rotating at the laser frequency):
\begin{align}
\notag	\hat{H}&=-\Delta\hat{a}^\dagger_1\hat{a}_1-(\Delta-\delta\omega)\hat{a}^\dagger_2\hat{a}_2+\Omega\hat{b}^\dagger\hat{b}\\
		&+g(\hat{a}^\dagger_2\hat{b}\hat{a}_1+\hat{a}_2\hat{b}^\dagger\hat{a}^\dagger_1)+\hat{H}^{laser}_{drive}+\hat{H}_{diss}~.
\end{align}
This leads to the following classical equations for $\langle\hat{a_1}\rangle=\alpha_1$ etc.:
\begin{align}
	\dot{\alpha_1}&=\left[i\Delta-\kappa_1/2\right]\alpha_1+\kappa_1/2\alpha^{max}_1-ig\alpha_2\beta^*~,\\
	\dot{\alpha_2}&=\left[i\left(\Delta-\delta\omega\right)-\kappa_2/2\right]\alpha_2-ig\alpha_1\beta~,\\
	\dot{\beta}&=\left[-i\Omega-\Gamma/2\right]\beta+\sqrt{n_{th}\Gamma}\xi(t)-ig\alpha_2\alpha^*_1~,
\end{align}
with $\langle\xi^*(t)\xi(t')\rangle=\delta(t-t')$. Here, $\alpha^{max}_1$ is the amplitude of the laser-driven mode at resonance ($\Delta=0$) in the absence of optomechanical coupling ($g=0$). $\delta\omega$ is the frequency difference between the optical modes. We solve this system analytically in the simplified case where the laser-drive mode is on resonance, and the frequency difference between modes is chosen to be equal to the mechanical mode frequency $\Omega$. Additionally, we assume a non-depleted pump and linearize the solution around $(\alpha_1,\alpha_2,\beta)=(\alpha^{max}_1,0,0)$. Taking these simplifications into account, the equations to be solved reduce to:
\begin{align}
	\label{a2start}
	\dot{\alpha_2}&=\left[-i\delta\omega-\kappa_2/2\right]\alpha_2-ig\alpha^{max}_1\beta~,\\ 
	\label{bstart}
	\dot{\beta}&=\left[i\Omega-\Gamma/2\right]\beta+\sqrt{n_{th}\Gamma}\xi(t)-ig\alpha_2\alpha^{max*}_1~.
\end{align}
We solve the system of equations in Fourier space to obtain an expression for $\langle|\beta(t)|^2\rangle = \bar{n}$ and $\langle|\alpha_2(t)|^2\rangle=\bar{n}_{aS}$, the average phonon number and anti-Stokes photon occupation respectively:
\begin{align}
	\langle|\beta(t)|^2\rangle&=-n_{th}\Gamma\\
		&\left(\frac{|c_1|^2}{2Re[s_1]}+\frac{(c_1 c_2^*)}{s_1+s_2^*}+\frac{(c_1^* c_2)}{s_1^*+s_2}+\frac{|c_2|^2}{2Re[s_2]}\right) \notag\\
	\langle|\alpha_2(t)|^2\rangle&=-g_{k,q}^2|\alpha^{max}_1|^2n_{th}\Gamma\\
		&\left(\frac{|c_3|^2}{2Re[s_1]}+\frac{(c_3 c_4^*)}{(s_1+s_2^*)}+\frac{(c_3^* c_4)}{(s_1^*+s_2)}+\frac{|c_4|^2}{2Re[s_2]}\right) \notag\\
		s_{1,2}=-&\left(\frac{\Gamma}{4}+\frac{\kappa}{4}\right)\pm\sqrt{\left(\frac{\Gamma}{4}-\frac{\kappa}{4}\right)^2-g^2|\alpha^{max}_1|^2}\\
%\end{align}
%
%\begin{align}
	c_1&=\frac{s_1+\frac{\kappa}{2}}{s_1-s_2},~~~~~~~~~c_2=\frac{s_2+\frac{\kappa}{2}}{s_2-s_1}\\
	c_3 &= 1/(s_1-s_2),~~~~c_4 = 1/(s_2-s_1)
\end{align}
%
%
%%%%%%%%%%%%%%%
%QUANTUM NOISE APPROACH%
%%%%%%%%%%%%%%%

\emph{Quantum noise approach.---}
Alternatively, we can employ the quantum noise approach to derive the cooling and heating rates for the mechanical mode $q$ subject to a situation where the mode $k$ is assumed to be laser-driven and photons are scattered into the other mode $k+q$. In contrast to the amplitude equations discussed above, the rate equation approach will only work for $\kappa>>\Gamma_{opt}+\Gamma_M$, but unlike the amplitude approach it is not limited to situations where the optical mode populations remain essentially unchanged by the Brillouin processes. The idea will be to replace the driven photon mode operator by a c-number and to view the resulting Hamiltonian as composed of a fluctuating quantum noise term coupling to the mechanical mode. This quantum noise term essentially arises from the interference between the driven mode and the vacuum fluctuations in the second mode. Consider the mode $k$ to be driven as $\hat{a}_k=\alpha e^{-i\omega_Lt}$. Upon substitution, the coupling term becomes:
    \begin{align}
    \hbar g_{k,q}(\hat{b}_q\hat{a}_{k+q}\alpha e^{-i\omega_Lt}+h.c.)~.
    \end{align}
where $\hat{b}_q$ couples to fluctuating quantum noise variable, $\hat{F}=\hbar g_{k,q}\hat{a}^\dagger_{k+q}\alpha e^{-i\omega_Lt}$, in the form $\hat{H}_{int}=\hat{b}_q\hat{F}^\dagger+\hat{b}^\dagger_q\hat{F}$. The transition rate for phonon annihilation is:
    \begin{align}
    \label{eqRateUp}
    \Gamma_{n-1\leftarrow n}&=n\frac{1}{\hbar^2}\left\langle\hat{F}\hat{F}^\dagger\right\rangle_{\omega=\Omega}~,\\
    \left\langle\hat{F}\hat{F}^\dagger\right\rangle_\omega&=(\hbar g_{k+q})^2\nb_{phot}2Re\left[\frac{-1}{i(\Delta+\omega)-\kappa/2}\right]~.
    \end{align}
Likewise, the transition rate for phonon creation is:
    \begin{align}
    \label{eqRateDown}
    \Gamma_{n\leftarrow n-1}=n\frac{1}{\hbar^2}\left\langle\hat{F}^\dagger\hat{F}\right\rangle_{\omega=-\Omega}~,\\
    \left\langle\hat{F}^\dagger\hat{F}\right\rangle_\omega=\int dt e^{-i\omega t}\left\langle\hat{F}^\dagger(t)\hat{F}(0)\right\rangle=0~.
    \end{align}
Taking into account both transition rates, we are left with a cooling rate that is set by the balance between up- and downward transitions. It can be written in the form:
   \begin{align}
   \Gamma_{opt}&=\Gamma_0 \bar{n}_{phot}~,\\
    \Gamma_{opt}&=g^2_{k+q}\frac{\kappa}{(\kappa/2)^2+(\omega_{k+q}-\omega_L-\Omega)^2}~,
    \end{align}
where $\Gamma_{opt}$ quantifies the rate of Brillouin scattering. Here we have split off the dependence on the photon number $\bar{n}_{phot}$ circulating inside the lower optical mode. Below, we will display the slightly generalized expressions for the average up- and down-transition rates (deduced from Eqs. \eqref{eqRateUp} and \eqref{eqRateDown}), for the case of arbitrary photon numbers in both optical modes.
%
%%%%%%%%%%%%%%%%
% Rate Equations %
%%%%%%%%%%

\emph{Rate equation approach.---} We can now proceed to solve a system of rate equations to determine the average phonon number in our system. We note that these rate equations do not take into account non-resonant scattering processes (scattering into the tails of the optical density of states, suppressed by a factor $(\kappa/\Omega)^2)$, and would also cease to be valid in a strong coupling regime (where $\Gamma_{opt}>\kappa,\Omega$). Having derived the form of $\Gamma_{opt}$, let us now consider a two level optical system with photon decay rates $\kappa_1$ and $\kappa_2$. The transition rates between the two levels will be written in terms of the optomechanical coupling rate $\Gamma_0$, the average photon occupations of the two optical states $\nb_{p},\nb_{aS}$, and the average phonon occupation $\nb$. The up transition corresponds to cooling and the down transition to heating,
    \begin{align}
    \Gu&=\Gamma_0\nb_{p}(\nb_{aS}+1)\nb~,\\
    \Gd&=\Gamma_0(\nb_{p}+1)\nb_{aS}(\nb+1)~.
    \end{align}
We can write the rate equations for the two levels based on the cavity decay rates and the heating and cooling transition rates, and the steady state photon number due to the driving laser $\bar{n}^L_{p}$,
    \begin{align}
    \nbd_{aS}&=-\nb_{aS}\kappa_2-\Gd+\Gu~,\\
    \nbd_{p}&=(\nb^L_{p}-\nb_{p})\kappa_1+\Gd-\Gu~,\\
    \nbd&=(\nb_{th}-\nb)\Gamma_M+\Gd-\Gu~.
    \end{align}
Solving for the steady state solution by setting the time derivatives equal to zero yields a set of equations which relate the photon occupations to the pump laser and acoustical phonon occupation:
    \begin{align}
    \nb_{p}&=\nb^L_{p}-(\Gamma/\kappa_1)~,\\
    \nb_{aS}&=(\Gamma/\kappa_2)~,\\
    \Gamma&\equiv\Gu-\Gd=(\nb_{th}-\nb)\Gamma_M~.
    \end{align}
We see that $\Gamma=\Gamma[\nb,\nb_{p}(\nb),\nb_{aS}(\nb)]$ is a nonlinear function of $\nb$. Thus we must solve the following relation,
    \begin{align}
    \Gamma(\nb)=(\nb_{th}-\nb)\Gamma_M~.
    \end{align}
where $\nb$ will be a function of the dimensionless parameters $\nb_{th},\nb^L_{p},\frac{\kappa_1}{\Gamma_M},\frac{\kappa_2}{\Gamma_M},\frac{\Gamma_0}{\Gamma_M}$.

\label{secExp}
\emph{Discussion.---}In the simple limit where the optical relaxtion rates $\kappa_1,\kappa_2$ are fast compared to the mechanical relaxation rate and optomechanical coupling rate $\Gamma_0,\Gamma_M$, we find only the cooling rate remains. In this case, the system simplifies to the following relations:
    \begin{align}
    &\Gd=0~, &\Gu&=\Gamma_0\nb^L_{p}\nb~,\\
    &\nb_{p}\approx\nb^L_{p}~, &\nb_{aS}&\approx0~,\\
    &\Gamma(\nb)=\Gu-\Gd=\Gamma_0\nb\nb^L_{p}~,\\
    &\nb=\nb_{th}\frac{\Gamma_M}{\Gamma_M+\Gamma_0\nb^L_{p}}~.
    \end{align}

This is the usual classical cooling result where the cooling is unlimited with input power. We can also solve the full set of equations numerically, and obtain the cooling rate as a function of input power. For the system, we take for example a 100$\mu$m diameter $SiO_2$ sphere\cite{Carmon2007b, xBahl2011b} such as the ones used in the context of Brillouin scattering previously\cite{Tomes2009,Bahl2011}. The acoustical mode is taken to be a 50 MHz surface type mode like what was theoretically suggested \cite{Matsko2009}, numerically calculated \cite{Zehnpfennig2011}, and experimentally observed \cite{Bahl2011} to be excited via forward Brillouin scattering \cite{Shelby1985}. The sphere is taken to have two optical modes of high quality factor $Q=4{\times}10^8$. The optical modes are exactly separated by the mechanical oscillation frequency. Our pump is a telecom compatible source ($\lambda=1.55 \mu m$). The relevant area is taken to be where the modes overlap, which is proportional to $\lambda^2$ (the optical mode area) \cite{Oxborrow2007} as the optical mode is much smaller than the acoustical one \cite{Zehnpfennig2011}.
\begin{figure}[tbp]
    \centering
    \includegraphics[width=.8\hsize]{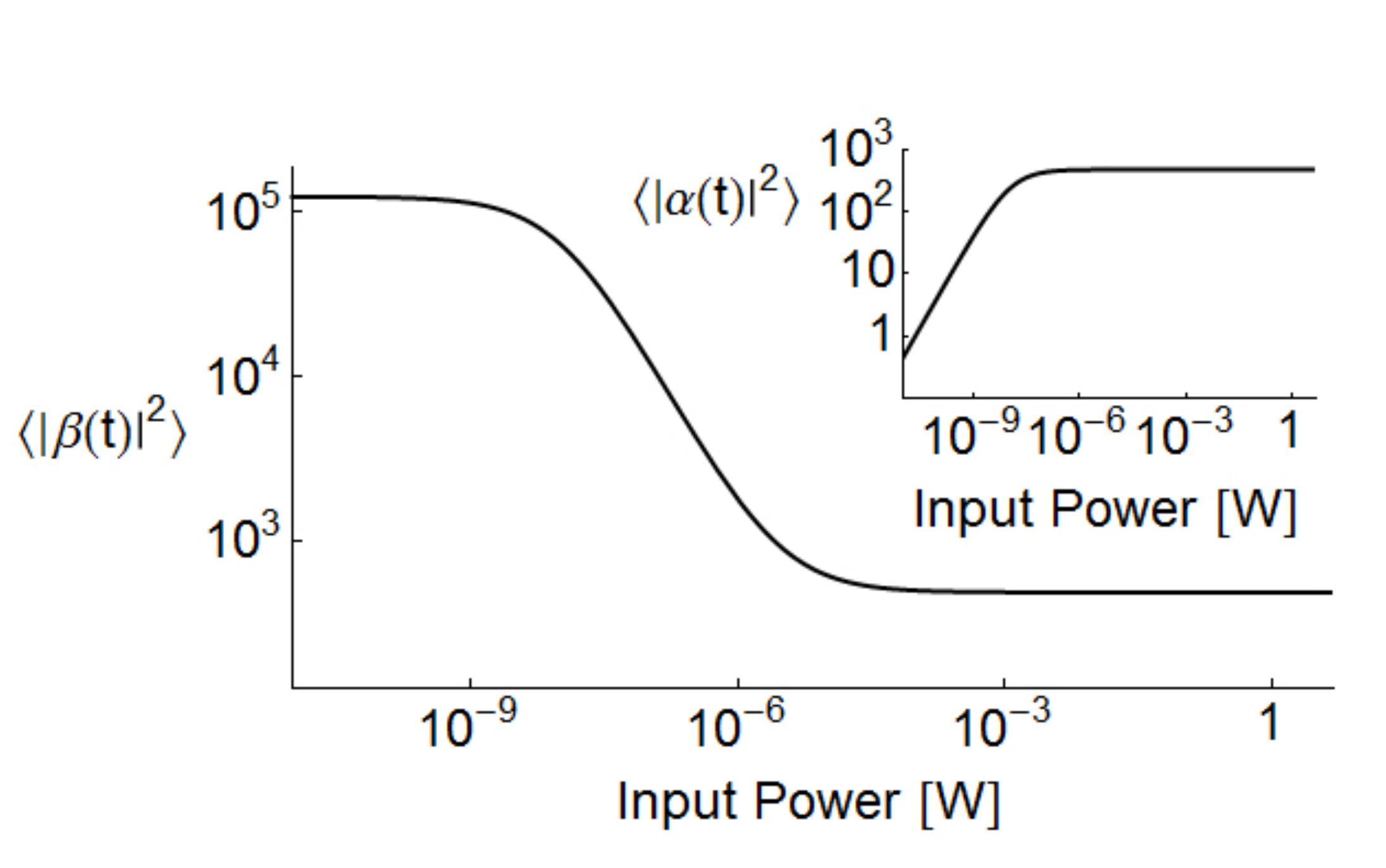}
    \caption{\label{fignvP}
    Average phonon number as a function of input power. The average phonon number starts from a thermal occupation of about $10^5$ and cools by a ratio of about 260. Inset: Intra-cavity anti-Stokes dependence on input power.}
\end{figure}
\begin{figure}[tbp]
    \centering
    \includegraphics[width=.9\hsize]{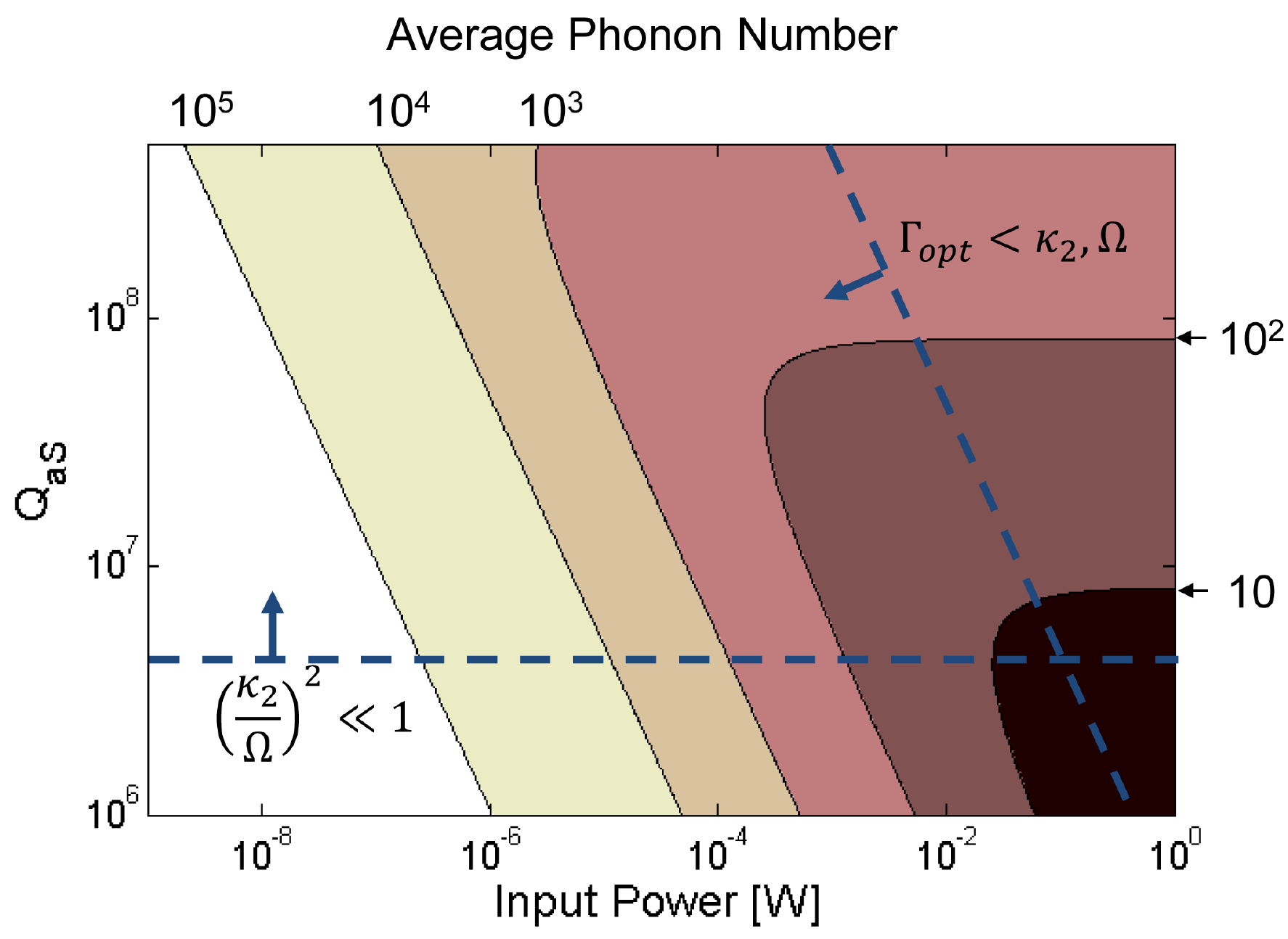}
    \caption{\label{nvPQ}
    Average phonon number as a function of both input power and anti-Stokes quality factor. The dashed lines indicate the regime where our approximations hold.}
\end{figure}

Figure \ref{fignvP} plots our theoretical prediction for the average phonon number in such a system as a function of input power. Significant cooling would begin at pump input powers of a few $nW$ and saturates after a few $mW$.  Starting from a room temperature phonon number of about $10^5$ (as is typical for such systems), one could cool by a ratio of about 260 in this example. As seen in the inset of Fig. \ref{fignvP}, the amount of power circulating in the anti-Stokes mode clamps as the cooling process begins. As input power is increased, the final average phonon number asymptotically converges to a lower limit:
    \begin{align}\label{eqnfinal}
    n_{final}=\frac{n_{th}}{\frac{\kappa_2}{\Gamma_M}+1}.
    \end{align}
Eq. \eqref{eqnfinal} indicates that if the quality factor of the anti-Stokes resonance were deliberately lowered compared to the pump resonance, higher cooling ratios could be acheived: Brillouin cooling requires an efficient way to get rid of the anti-Stokes photons. We therefore determine the average phonon value as a function of both pump power and anti-Stokes mode quality factor, while all other parameters are left unchanged. Though many experimental challenges need to be overcome, as seen in Fig. \ref{nvPQ}, for diminished anti-Stokes quality factors, cooling ratios above $10^4$, can be acheived.

\emph{Conclusion.---}We describe here a triply resonant structure for Brillouin cooling and develop the theory describing this system. 
Unlike before \cite{Metzger2004,Kleckner2006,Gigan2006,Carmon2005,Kippenberg2005,Rokhsari2005,Eichenfield2009}, for Brillouin cooling, the Doppler reflector is a monotonically travelling acoustic wave. Additionally, different from fluorescent anti-Stokes cooling \cite{Epstein2009} that cools the whole thermal bath, Brillouin cooling evacuates heat from one selected vibrational mode, making it attractive for experiments in which this particular natural frequency is addressed \cite{Mancini2002,Marshall2003}.

\begin{acknowledgments}
This work was supported by the Defense Advanced Research Projects Agency (DARPA) Optical Radiation Cooling and Heating in Integrated Devices (ORCHID) program through a grant from the Air Force Office of Scientific Research (AFOSR). M.T. is supported by a National Science Foundation fellowship. F.M. acknowledges the Emmy-Noether program.
\end{acknowledgments}

%\bibliography{BrillouinCooling,BrillouinCoolingSpecial}
%\bibliographystyle{ieeetr}

%\input{brillouin_cooling.bbl}

\end{document}